\documentclass[seceq]{ptptex}

\usepackage{graphicx}

\title{
A brief overview of fixed-order  perturbative QCD calculations \\ of jet production 
in heavy-ion collisions%
}

\author{
Ivan \textsc{Vitev}%
}

\inst{
Los Alamos National Laboratory \\ Theoretical Division, Group T-2, 
Mail Stop B283, Los Alamos, NM 87545, USA

}

\abst{

We review recent  developments in the QCD description of jet production and 
modification in reactions with heavy nuclei at relativistic energies. Our goal is to 
formulate a perturbative expansion in the presence of  nuclear matter that
allows to systematically improve the accuracy of the theoretical predictions.   
As an example, we present calculations of inclusive jet cross sections at RHIC, 
 $Z^0/\gamma^*$-tagged jet cross sections at the LHC, and jet shapes 
that include both next-to-leading order perturbative effects and the 
effects of the nuclear medium.  

}

\begin{document}

\maketitle

\section{Introduction}

In the past several years important developments in jet finding algorithms~\cite{JetFind}  combined with  
advances in detector technology and experimental analysis have enabled, for the first time, measurements 
of jets~\cite{Sterman:1977wj} in nuclear reactions at very high energies at the Relativistic Heavy Ion Collider 
(RHIC)~\cite{JetExp}. Observation of jets  is  among the first physics results reported at the Large Hadron Collider 
(LHC)~\cite{mikko}, which also has an active heavy-ion program~\cite{jetlhc}. In nucleus-nucleus (A+A) collisions, 
jet  observables are expected to be much  more discriminating~\cite{VWZ} 
with respect to the underlying modes of parton propagation and energy  loss in strongly-interacting 
matter~\cite{models} than existing studies of leading particles and leading particle correlations.
Developments if theory are on the way to guide the heavy-ion experimental jet programs at RHIC and at the LHC and to
help interpret current and upcoming physics results.

To take full advantage of jet physics, calculations at next-to-leading  order (NLO) and beyond in perturbative 
Quantum Chromodynamics (QCD) are required~\cite{EKS,MCFM}. In heavy-ion reactions there is the added complication 
of a soft background  medium  that these jets must traverse - the quark-gluon plasma (QGP). There are multiple ways to 
describe its properties  that include but are not limited to temperature $T$, 
coupling strength $g^{\rm med.}$, density $\rho$, energy density $\epsilon$, parton rapidity density $dN^g/dy$, 
Debye screening scale $m_D$, parton mean free paths $\lambda_{q}$, $\lambda_{g}$, transport coefficient $\hat{q}$, 
and the strength of  the background gluon field $\langle F^{+\perp} F_{\perp}^+  \rangle$. 
Relations between some of these quantities  can be derived for specific models of the nuclear matter. 
In practice, however, one always needs independent 
constrains from experimental data or lattice simulations since the QGP that A+A reactions aim to produce 
may be strongly-coupled or non-perturbative. 

The production of jets and the medium-induced
bremsstrahlung at scales $Q^2 \sim E_T^2 \gg \Lambda_{QCD}^2$ and $Q^2 \sim (gT)E_T \gg   \Lambda_{QCD}^2 $,
on the other hand, 
can be treated perturbatively.  In fact, parton energy loss processes in the  QGP and one-loop perturbative 
corrections are formally manifested  in experimental observables at the same order 
${\cal O}(\alpha_s^3)$, ${\cal O}( \alpha_s^2\alpha_s^{\rm med.})$. It should be noted, however, that the values of 
$\alpha_s$ and $\alpha_s^{\rm med.}$ can differ considerably. The emerging perturbative expansion for jet 
 production and modification in heavy-ion reactions can easily be generalized to higher orders and is shown
in Table~\ref{orders}. Self-consistent fixed-order theoretical calculations are the ones that take into consideration 
all terms along the counter diagonals of this matrix. At present, only inclusive $\alpha_s^{\rm med.}$ induced gluon 
bremsstrahlung  has been evaluated and the corresponding next-to-leading order jet calculations  are presented below.

\begin{table}[!t]
\centering 
\begin{tabular}{c c c c c c } 
\hline\hline 
 \hspace*{.3cm}Medium $\rightarrow$   & LO & NLO & NNLO & NNNLO & \ \\ [0.5ex] 
Vacuum $\downarrow$    & & & & \\  
\hline   
  LO &  $\alpha_s^2$  & $\alpha_s^2 (\alpha_s^{\rm med.})$  & $\alpha_s^2 (\alpha_s^{\rm med.})^2$ 
& $\alpha_s^2 (\alpha_s^{\rm med.})^3$ & $\cdots$\\[.5ex] 
 NLO & $\alpha_s^3$  & $\alpha_s^3 (\alpha_s^{\rm med.}) $ & $\alpha_s^3 (\alpha_s^{\rm med.})^2$ 
& $\alpha_s^3 (\alpha_s^{\rm med.})^3 $& $\cdots$\\[.5ex] 
 NNLO &  $\alpha_s^4$  & $\alpha_s^4 (\alpha_s^{\rm med.}) $ & $\alpha_s^4 (\alpha_s^{\rm med.})^2$ 
& $\alpha_s^4 (\alpha_s^{\rm med.})^3$ & $\cdots$\\[.5ex] 
\hline 
\end{tabular}
\caption{Perturbative expansion relevant to large $Q^2$ processes in the presence of a nuclear
medium. }   
\label{orders} 
\end{table}

\section{Inclusive jet production at RHIC}

\begin{figure}[!t]
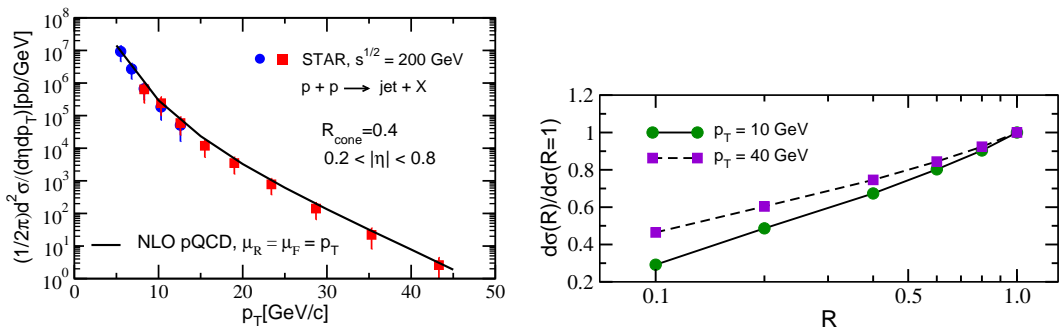

\vspace*{-3cm}
\includegraphics[width=2.6in,angle=0]{STAR_pp_comp.eps}\hspace*{.3cm}
\includegraphics[width=2.7in,angle=0]{STAR-Rcone.eps}
\caption{Comparison of the NLO calculation to STAR experimental data~\cite{Abelev:2006uq} 
on the inclusive jet cross section for R=0.4 (left panel). The variation of the jet cross 
section with the cone size $R$ for  $E_T=10,\,40$~GeV around midrapidity at RHIC 
is also shown (right panel).}
\label{pp}
\end{figure}

At NLO, the  inclusive and tagged jets cross sections can be expressed schematically as follows:
\begin{eqnarray}
d\sigma_{jet (+{\rm tag})}& =&\frac{1}{2!}
d\sigma[2\rightarrow 2] S_2 (\{p,y,\phi\}_2) + \frac{1}{3!}
   d\sigma[2\rightarrow 3] S_3 (\{p,y,\phi\}_3 ) \; .
\label{eq:CS_NLO}
\end{eqnarray}
Here, $p_i,y_i,\phi_i$ are the transverse momentum, rapidity, and azimuthal angle
of the i-th particle ($i=1,2,3$), respectively, and
$\sigma[2\rightarrow 2]$, $\sigma[2\rightarrow 3]$ represent the production cross sections 
with two and three final-state partons.  $S_2$ and $S_3$ are phase space constraints  
and  $S_2 = \sum_{i=1}^{2} S(i)= 
\sum_{i=1}^{2} \delta(E_{T_i} - E_T)\delta(y_i -y)$  identifies the jet with its parent parton. Hence,
it is only at next-to-leading order that the dependence of the experimental 
observables on the jet cone radius $R$, the jet finding 
algorithm, or the trigger particle energy can be theoretically investigated.  
For an angular  separation $R_{ij}=\sqrt{(y_i - y_j)^2 + (\phi_i - \phi_j)^2}$, defined for any
possible parton pair $(i,j)$,
\begin{eqnarray}
\label{eq:2to3}
&&S_3 = \sum_{i<j} \delta(E_{T_i}+E_{T_j}-E_{T})
\delta\left(\frac{E_{T_i} y_i + E_{T_j} y_j}{E_{T_i}+E_{T_j}} - y\right)\theta\left(R_{ij} < R_{\rm rc} \right) \\
&&
\hspace*{0cm}+  \sum_i S(i)\prod_{j\neq i} 
\theta\left(R_{ij} > 
\frac{(E_{T_i}+E_{T_j})R}{\max(E_{T_i},E_{T_j})} \right) \,,  
R_{\rm rc} = \min\left(R_{sep}R, 
\frac{E_{T_i}+E_{T_j}}{\max(E_{T_i},E_{T_j})} R \right)\,. \nonumber 
\end{eqnarray}
In Eq.~(\ref{eq:2to3}) $R_{\rm rc}$  determines when two partons 
should be  recombined in a jet. Here, $1 \leq R_{sep} \leq 2$ is introduced 
to take into account features of experimental cone algorithms, employed to improve 
infrared safety. Eq.~(\ref{eq:2to3}) establishes a correspondence between the
commonly used jet finders and the perturbative calculations 
to ${\cal O}(\alpha_s^3)$ with the goal of providing accurate predictions for  
comparison to data. For example,  $R_{sep} = 2$ yields a midpoint cone algorithm and  
$R_{sep} = 1$ corresponds to the 
$k_T$   algorithm~\cite{EKS,Seymour:1997kj}. 
$R$ is the cone size or parton   separation parameter, respectively.
We have compared  NLO calculation~\cite{EKS} of the inclusive jet cross section at ${\sqrt s}=200$~GeV p+p 
collisions  at RHIC to the STAR experimental measurement which uses a midpoint cone
algorithm~\cite{Abelev:2006uq} in the pseudorapidity range $ 0.2 \leq \eta \leq 0.8 $. 
Very good agreement between data and theory is achieved with a standard 
choice for the renormalization and factorization scales  
$\mu_R =\mu_f =E_T$~\cite{Vitev:2009rd}, as shown in the left panel of Figure~\ref{pp}. 
Variation of these scales within $(E_T/2,2E_T)$ leads to
less than ($+10\% , -20\%$) variation of the jet cross section. We also found
that there is a   significant dependence of 
$d\sigma^{\rm jet}/dydE_T$  on the cone size $R$, which, even in p+p reactions,
can exceed a factor of two. This is illustrated in the right panel of Figure~\ref{pp}.   
Analytically, the $\ln (R/R_0)$ scaling of the 
cross section can be understood from the $1/r$ angular behavior of the perturbative QCD 
splitting kernel at high energies.

\begin{figure}[!t]
\includegraphics[width=2.6in,angle=0]{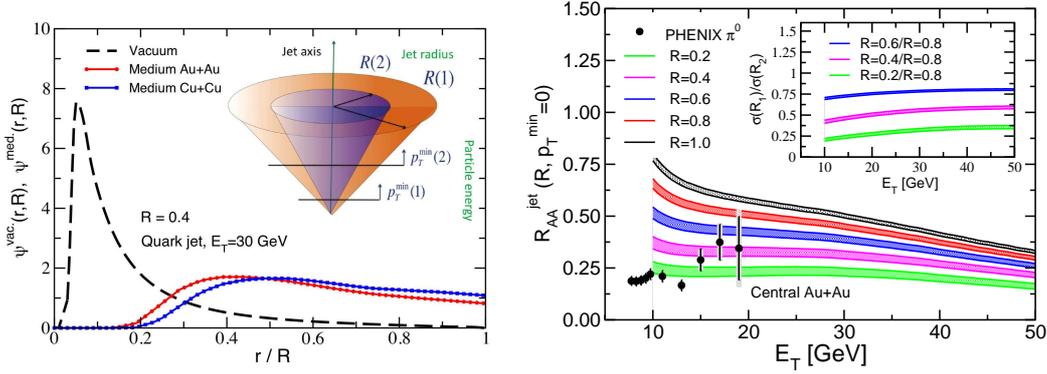}\hspace*{.3cm}
\includegraphics[width=2.7in,angle=0]{Raa-R-b3-jetAu-cold-IV.eps}
\caption{The differential jet shape in vacuum $\psi^{\rm vac.}(r,R)$ is contrasted to the  medium-induced 
contribution  $\psi^{\rm med.}(r,R)$  by a  $E_T = 30$~GeV quark in Au+Au and Cu+Cu collisions 
at $\sqrt{s_{NN}}=200$~GeV. The insert illustrates a method for studying the
characteristics of these parton showers (left panel).  
Transverse energy dependent nuclear modification factor $R_{AA}^{\rm jet}$
for different cone radii $R$ in $b=3$~fm Au+Au  collisions.  Inserts show ratios of jet cross 
sections for different $R$ in nuclear reactions versus  $E_T$ (right panel). }
\label{illust}
\end{figure}

When compared to a parton shower in the vacuum, the medium-induced quark and gluon 
splittings have noticeably different angular and lightcone momentum fraction 
dependencies~\cite{VWZ,Vitev:2005yg}. In particular, for energetic partons propagating 
in hot and dense QCD matter, 
the origin of the coherent suppression of their radiative energy loss, known as the 
Landau-Pomeranchuk-Migdal effect, can be traced to the cancellation of the collinear 
radiation  at $r < m_D / \langle\, \omega (m_D,\lambda_g, E_T) \, 
\rangle$~\cite{Vitev:2005yg}.  Here, the  Debye screening 
scale $m_D  \simeq gT$ and $\langle \, \omega \, \rangle \simeq$~ few GeV. 
Thus, the medium-induced component of the jet, which is given by the properly normalized 
gluon bremsstrahlung intensity spectrum $ \psi^{\rm med}(r,R) \propto dI^{\rm rad}/d\omega dr $ 
within  the cone, has a characteristic large-angle distribution
away from the jet axis. This is illustrated in Figure~\ref{illust} for central Au+Au and 
central Cu+Cu collisions at RHIC. We emphasize that accurate numerical simulations, taking into 
account the geometry of the heavy-ion reaction, the longitudinal Bjorken expansion
of the QGP, and the constraints imposed by its experimentally measured 
entropy density per unit rapidity~\cite{VWZ}, have been performed for all physics results 
quoted here.

One can exploit the differences between the vacuum and the in-medium parton showers
by varying the cone radius $R$ ($R_{i,\rm jet} < R$) and a cut $p_T^{\min}$ ($E_{T_i} > p_T^{\min}$) 
for the particles "$i$'' that constitute the jet, to gain sensitivity to the properties  of 
the QGP and  of the mechanisms of parton energy loss in hot
and dense QCD matter~\cite{VWZ}. This is illustrated in the insert of Figure~\ref{illust}.  
The most easily accessible experimental feature of jet production in nuclear collisions 
is, arguably,  the suppression of the inclusive cross section in heavy-ion reaction 
compared to the binary collision scaled, $\propto \langle  N_{\rm bin}  \rangle$, 
production rate in elementary nucleon-nucleon reactions~\cite{Vitev:2009rd}: 
\begin{equation}
R_{AA}^{\text{jet}}(E_T; R,p_T^{\min}) =  \frac{d\sigma^{AA}(E_T;R,p_T^{\min})}{dy d^2 E_T} \Bigg/
 \langle  N_{\rm bin}  \rangle
\frac{ d\sigma^{pp}(E_T;R,p_T^{\min})}{dy d^2 E_T}   \; . 
\label{RAAjet}
\end{equation}
Eq.~(\ref{RAAjet}) defines a  two dimensional jet attenuation pattern 
versus $R$ and $p_T^{\min}$ for every fixed $E_T$. In contrast, for the same $E_T$, 
inclusive  particle quenching is represented by a  single 
value  related to the $R \rightarrow 0$ and
$p_T^{\min} \gg \langle \, \omega \, \rangle $ limit in Eq.~(\ref{RAAjet}).
Thus, jet observables are much more differential and, hence, 
immensely more powerful than leading particles and leading particle correlations 
in their ability to discriminate between the competing  
physics mechanisms of quark and gluon energy loss in dense QCD matter and between 
theoretical model approximations to parton dynamics in the QGP.

We calculate the medium-modified jet cross section per
binary nucleon-nucleon scattering as follows ($p_T^{\min}=0$):
\begin{equation}
\frac{1}{\langle  N_{\rm bin}  \rangle} 
\frac{d\sigma^{AA}(R)}{dyd^2E_T} = \int_{\epsilon=0}^1
d\epsilon \; \sum_{q,g}  P_{q,g}(\epsilon,E) 
\frac{1}{ (1 - (1-f_{q,g}) \cdot \epsilon)^2} 
 \frac{d\sigma^{\rm CNM,NLO}_{q,g}(R)} {dyd^2E^\prime_T} \; . 
\label{eq:JCS-AA}
\end{equation}
Here, $P_{q,g}(\epsilon,E)$ is the probability distribution for the parent quarks and gluons
to lose a fraction $\epsilon=\sum_i \omega_i/E$ of their energy due to multiple gluon emission
in the QGP.
In Eq.~(\ref{eq:JCS-AA})   ${d\sigma^{\rm CNM,NLO}_{q,g}(R)} /{dyd^2E^\prime_T}$ is the differential 
cross section which includes the known cold nuclear matter effects~\cite{CNM} and 
$(1-f_{q,g}) \cdot \epsilon$  represents the fraction of the 
energy of the parent parton that the medium re-distributes outside of the cone of
radius $R$. The measured cross section is then a probabilistic superposition of 
the  cross sections of protojets of initially larger energy
$E^\prime_T = E_T / (1 - (1-f_{q,g})\cdot \epsilon)$. Our results for
the nuclear modification factor of inclusive jets $R_{AA}^{\rm jet}$ in central Au+Au
collisions with $\sqrt{s_{NN}}=200$~GeV at RHIC are presented 
in the right panel of Figure~\ref{illust}.  Experimental data on leading $\pi^0$ suppression for these
reactions is only included for reference. 
A continuous variation of $R_{AA}^{\rm jet}$ with the cone radius $R$ is 
clearly observed  and shows the sensitivity of the inclusive 
jet cross section  in high-energy nuclear collisions to the characteristics of
QGP-induced parton shower. For $R \leq 0.2$ the quenching of jets approximates the
already observed suppression in the production  rate of inclusive high-$p_T$ particles.
It should be noted that in our theoretical calculation CNM effects contribute close 
to $1/2$ of the observed attenuation for $E_T  \geq 30$~GeV. These can be 
dramatically reduced at all $E_T$ by taking the ratio of two differential cross 
section  measurements for different cone radii $R_1$ and $R_2$, as shown in the insert.

\section{$Z^0/\gamma^*$-tagged jets at  the LHC}

We begin by discussing the cross section for $Z^0/\gamma^*$-tagged jet production in p+p
collisions.  It is instructive to first consider the leading order (LO) result, from which
one can understand the underlying production processes and appreciate why the $Z^0$ boson
was originally considered as a suitable tag for the initial associated jet
energy~\cite{Srivastava:2002kg}.  In the collinear factorization approach, the $Z^0/\gamma^*$+jet
cross section reads:
\begin{equation}
\frac{d \sigma}{d y_{(Z)} \, d y_{\rm(jet)} \, d^2 {\bf p}_{T\,(Z)} d^2 {\bf p}_{T\,\rm (jet)} }  =
\sum_{g,q, {q}} 
  \frac{f( {x}_1,\mu)f( {x}_2,\mu)|M|^2}{(2\pi)^2 \, 4 \,  {x}_1 \,  {x}_2 \, S^2}
\delta^2( {\bf p}_{T\,(Z)} - {\bf p}_{T\,\rm (jet)} )\; . 
\label{lotex}
\end{equation}
Here, $f(x_i,\mu)$ are
the parton distribution functions and $|M|^2$
are the relevant squared matrix elements.  The $\delta$-function constraint on the
transverse momentum of the jet is valid only at this order and only at the partonic level.

\begin{figure}[!t]
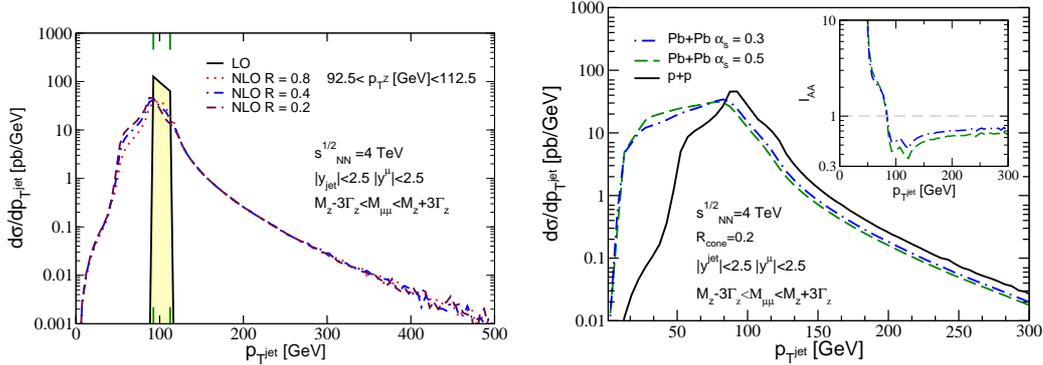

\includegraphics[width=2.6in]{925_1125_I.eps}\hspace*{.3cm}
\includegraphics[width=2.6in]{925_1125_r02.eps}
\caption{Transverse momentum distributions of a jet associated with 
$Z^0/\gamma^*(\rightarrow \mu^++\mu^-)$ tag to ${\cal O}(G_F\alpha_s)$ and ${\cal O}(G_F\alpha_s^2)$
for transverse momentum cut $92.5\;{\rm GeV} < p_T < 112.5$~GeV  on the tagging particle (left panel). 
The NLO $p_T$-differential cross section per nucleon pair for these tagged jets 
in central Pb+Pb collisions at the LHC is also presented for a cone radius $R=0.2$ (right panel). 
The ratio of the two cross sections is shown in the insert.}
\label{z0}
\end{figure}

The main advantage of the next-to-leading order $Z^0/\gamma^*$+jet+X calculation~\cite{MCFM} is the ability to
precisely predict the transverse momentum distribution of jets associated with a dimuon tag in a narrow
$p_T$ interval~\cite{Neufeld:2010fj}. Beyond tree level, the momentum constraint that the $Z^0$ boson 
measurement provides
is compromised by parton splitting and Z-strahlung processes.  We demonstrate this in 
the left panel of Figure~\ref{z0},
which shows the differential cross section for jets tagged with
$Z^0$/$\gamma^*\rightarrow \mu^+\mu^-$ in p+p collisions at $\sqrt{s_{NN}} = 4$ TeV.  
We implement acceptance cuts of $|y| < 2.5$ for both jets and
final-state muons, and, as mentioned above, constrain the invariant mass of the dimuon pair to the
interval $M_z \pm 3\Gamma_z$, where $M_z = 91.1876$ GeV and $\Gamma_z = 2.4952$ GeV.  This is the kinematic
acceptance range in which we will evaluate all results that follow and which can be easily adjusted
to match upcoming experimental measurements.
For the cross section shown in Figure~\ref{z0}, the tagging $Z^0$/$\gamma^*$ is required to
have $92.5\; {\rm GeV} < p_T < 112.5$~GeV.
The LO result restricts the $p_T$ of the jet to lie exactly within this interval, consistent with Eq.~(\ref{lotex}).
As seen in the left panel of Figure~\ref{z0}, at NLO  the  deviations from this naive relation are very significant.
We have included results for three different values of the jet cone radius, $R=0.2, \; 0.4, 0.8$. The
variation of the cross section with $R$ around $p_{T\, \rm (jet)} \sim p_{T\,(Z)}$ arises from the  interplay
between the amount of energy that is contained in the jet and the number of reconstructed jets (one or two).
For $p_{T\, \rm (jet)}\gg p_{T(Z)}$ or $p_{T\, \rm(jet)}\ll p_{T\, (Z)}$ the two final-state partons are
well-separated and identified as different jets. The falloff of the differential cross section relative
to its peak value at $p_{T\, \rm (jet)} = p_{T\,(Z)}$ is then controlled by the QCD splitting kernel
(the part related to the large lightcone parton momentum)  and there is no dependence on the cone radius.

In order to quantify the inability of the $Z^0/\gamma^*$ tag to constrain the momentum of the jet
we calculate the mean $p_T \equiv \langle p_{T\, \rm(jet)}\rangle$ and standard deviation
$\Delta p_{T\, \rm(jet)}  = \sqrt{\langle p_{T\, \rm(jet)}^2 \rangle - \langle p_{T\, \rm(jet)} \rangle^2}$
for each of the curves in Figure~\ref{z0}.  Our results are presented in Table~2.
The standard deviation for the LO curves is not strictly  zero because of the finite $p_T$ width of
the tagging $Z^0$/$\gamma^*$ interval. The NLO curves exhibit a similar $\langle p_{T\, \rm(jet)}\rangle$ as
the LO result, with  $\langle p_{T\, \rm(jet)} \rangle$
increasing as the cone radius increases. However, in going from LO to NLO there is a significant jump
in $\Delta p_{T\, \rm(jet)}$. The width of the jet momentum distribution  quadruples for an
energetic tag. The very large values of
$ \Delta p_{T\, \rm(jet)} / \langle p_{T\, \rm(jet)} \rangle   \sim 25 \%$ at NLO create serious
complications for experimentally tagging the initial associated jet energy in both p+p and A+A collisions.
While additional cuts can be considered, such as the requirement for a single jet within the
experimental acceptance that is exactly opposite the tagging particle in azimuth, these will
reduce the already small projected multiplicity for this final state in heavy-ion reactions.

\begin{table}[!b]
\begin{center} 
\label{devs}
\begin{tabular}{c c c c c c c} 
\hline\hline 
 $p_{T\, (Z)} $ [GeV] &    & LO & R = 0.2 & R = 0.4 & R = 0.8 \\ [0.5ex] 
\hline   
  92.5-112.5 &  \!\!\! $\langle p_{T\, \rm (jet)} \rangle$ [GeV] & 100.79 & 93.91 & 96.63  & 100.05\\[.5ex] 
&  \!\!\!\!\!\!\! $\Delta  p_{T\, \rm (jet)}$ [GeV] & 6.95 & 25.19 & 24.88 & 24.15\\ [1ex] 
\hline 
\end{tabular}
\label{devs}
\caption{Mean $p_T$ and $\Delta \langle p_T \rangle$ for $Z^0/\gamma^*$-tagged jets at the LHC from 
LO and NLO calculation.} 
\end{center}
\end{table}

Accounting for the fact that the $Z^0/\gamma^*$ and their decay dileptons escape the region of dense
nuclear matter unaffected by the strong interaction, the modified jet cross section can be calculated as  
follows:  
\begin{eqnarray}
\frac{d\sigma}{d^2 {\bf p}_{(Z)} d^2 {\bf p}_Q} &=& 
\sum_{q,g} \int d \epsilon \; 
\frac{P_{q,g}(\epsilon) }{[1 - (1-f_{q,g})\epsilon]^2} 
 \frac{d\sigma^{q,g}}{d^2 {\bf p}_{(Z)} d^2 {\bf p}_{\rm(jet)}}  \left(\frac{{\bf p}_{Q}}
{[1 - (1-f_{q,g})\epsilon)]}\right) \;. \qquad
\label{2Dquench}
\end{eqnarray}
Eq.~(\ref{2Dquench}) resembles the result for inclusive jet suppression and  we have integrated over $y_{\rm jet}$ and  $y_{Z}$
in finite rapidity intervals. Our results are presented in the right panel of Figure~\ref{z0},
where the tagging $Z^0$/$\gamma^*$ is required to have $92.5 \; {\rm GeV} < p_T < 112.5\; {\rm GeV} $.
The main physics effect that this figure illustrates is the QGP-induced modification to the 
vacuum parton shower. Specifically, its broadening \cite{Vitev:2005yg} implies that part of the jet 
energy is redistributed outside of the jet cone and the differential jet distribution is downshifted
toward smaller transverse momenta. The smaller the jet cone radius the more pronounced this effect is.
As the jet cone radius is increased, the medium-modified curves approach the p+p result, as more 
and more of the medium-induced bremsstrahlung is recovered in the jet.

A generalization of  the cross section ratios per binary nucleon-nucleon collisions for 
tagged jets~\cite{Neufeld:2010fj}, $I_{AA}^{\rm jet}$, is given in the insert in Figure~\ref{z0}.
The variation in the magnitude of $I_{AA}^{\rm jet}$ at transverse momenta larger than the $p_T$ of the tag 
is controlled by the shape of the jet spectrum.
The most striking feature is the sharp transition from tagged jet suppression above $p_{T\, (Z)}$  to tagged jet enhancement
below $p_{T\, (Z)}$. For the example shown in the left panel of Figure~\ref{z0}, the enhancement can be as large as a factor of ten.   
This transition from suppression to enhancement is a unique prediction of jet quenching for tagged jets and will constitute
an unambiguous experimental evidence for strong final-state interactions and parton energy loss in the QGP.

\section{Jet shapes at RHIC and at the LHC}

Integral and differential jet shapes, defined as: 
\begin{eqnarray}
&& \Psi_{\text{int}}(r,R) = \frac{\sum_i (E_T)_i \Theta (r-(R_{\text{jet}})_i)}
{\sum_i  (E_T)_i  \Theta(R-(R_{\text{jet}})_i)}\; , \quad
  \psi(r,R) = \frac{d\Psi_{\text{int}}(r,R)}{dr}  \;,
\end{eqnarray} 
were historically the first observables used to study jet sub-structure and intra-jet energy 
flow~\cite{Seymour:1997kj}. To ${\cal O}(\alpha_s^3)$  $\psi(r,R)$  can be
evaluated from the LO parton splitting functions in perturbative QCD~\cite{Seymour:1997kj} and 
the analytic results can be generalized  to allow for a minimum particle or calorimeter tower energy 
$p_T^{\min}$ in the definition of the jet~\cite{VWZ}. Such experimental cuts can be particularly useful 
in reducing the large background of soft particles in the high multiplicity environment of 
heavy-ion collisions.

Analytically, jet shapes are evaluated as follows~\cite{Seymour:1997kj,VWZ}:
\begin{eqnarray}
\label{totpsi}
\psi^{\rm vac.}(r,R)&=&\psi_{\text{coll}}(r,R)\left( P_{\rm Sudakov}(r,R)-1\right) +
\psi_{\text{LO}}(r,R) \\
&&  + \psi_{i,\text{LO}}(r,R)  + \psi_{\text{PC}}(r,R)
+ \psi_{i,\text{PC}}(r,R) \;.  \nonumber 
\end{eqnarray} 
In Eq.~(\ref{totpsi})  the first term represents  the  contribution from the Sudakov-resummed 
small-angle  parton splitting; the second and third terms give the leading-order   
final-state and initial-state  contributions, respectively; the last two 
terms come from power corrections $\propto Q_0/E_T$, $Q_0 \simeq 2-3$~GeV, when one integrates 
over the Landau pole in the modified leading logarithmic  approximation (MLLA). 
This approach was shown to provide a very good description of the differential 
intra-jet energy  flow at the Tevatron~\cite{VWZ}, as measured by CDF II~\cite{Acosta:2005ix}. 
Thus, reliable  predictions for the jet sub-structure in p+p reactions at RHIC and the LHC 
can be obtained  and used as a baseline to study the distortion of jet shapes in more complex 
systems, such as p+A and A+A.

Inclusive jet cross sections and jet shapes in nuclear collisions are closely 
related~\cite{VWZ,Vitev:2009rd}:
\begin{eqnarray}
  \label{psitotmed}
  \psi_{\rm tot.}\left(\frac{r}{R}\right) &=& \frac{ \langle  N_{\rm bin}  \rangle } { d\sigma^{AA}(R)/dyd^2E_T }
     \int_{\epsilon=0}^1
d\epsilon \; \sum_{q,g} \frac{P_{q,g}(\epsilon,E)  } 
{ (1 - (1-f_{q,g}) \cdot \epsilon)^3} \,   \\
&& \times   \frac{d\sigma^{\rm CNM,NLO}_{q,g}(R, E_T^\prime )}{dy d^2E_T^\prime}    \  \Bigg[ (1- \epsilon) \;
\psi_{\rm vac.}^{q,g}\left(\frac{r}{R};E^\prime \right) + f_{q,g}\cdot \epsilon \;
\psi_{\rm med.}^{q,g}\left(\frac{r}{R};E^\prime \right) \Bigg] \; .  \nonumber
\end{eqnarray}
It should be noted that 
vacuum and medium-induced parton showers become more collimated with increasing 
$E_T^\prime$ and the mean relative jet width,
\begin{equation}
 \left\langle \frac{r}{R}\right\rangle = \int_0^1 d\left(\frac{r}{R}\right)   \, \frac{r}{R} \,
\psi\left( \frac{r}{R}\right) \;,
\end{equation}
is reduced~\cite{VWZ,Vitev:2009rd}. Consequently,
the striking suppression pattern for jets, see for example the right panel of Figure~\ref{illust}, can 
be accompanied by a very modest growth in the observed 
$\langle r/R \rangle$.
We show in Table~\ref{table:mean-radii} the relative widths in the vacuum, for a
hypothetical case  of complete parton energy loss ($P_{q,g}(\epsilon)=\delta(1-\epsilon)$) 
that falls inside of $R$  ($f_{q,g}=1$), and for our realistic simulation 
of  $E_T = 20, \ 50, \ 100,$ and 200~GeV jets of $ R=0.4$ 
in central Pb+Pb collisions at the LHC.   These numerical results show that there is
very little $<10\%$ increase in the magnitude of this observable.
For  Au+Au and Cu+Cu collisions at RHIC,  we fond that, on average,  
jet broadening  is even smaller, $ <5\%$. 
Therefore, a rough 1-parameter characterization of energy flow in jets will not resolve the effect
of the QGP medium.

\begin{table}[!b]
\begin{center}
\begin{tabular}{c c c c }
\hline \hline
   $ R=0.4 $   & \ \  \ Vacuum \  & \  Complete E-loss \   & \ Realistic case \   \\[.5ex]
\hline
   $\langle r/R \rangle$, $E_T=20$GeV   &  0.41   & 0.55   &  0.45   \\[.5ex]
   $\langle r/R \rangle$, $E_T=50$GeV   &  0.35   & 0.48   &  0.38  \\[.9ex]
   $\langle r/R \rangle$, $E_T=100$GeV  &  0.28   & 0.44   &  0.32  \\[.9ex]
   $\langle r/R \rangle$, $E_T=200$GeV  &  0.25   & 0.40   &  0.28  \\[.9ex]
\hline 
\end{tabular}
\caption{ Summary of mean relative jet radii $\langle r/R \rangle$
in the vacuum, with complete energy loss, and  in the QGP medium.
Shown are results for cone radii $R=0.4$  and  transverse
energies $E_T = 20, 50, 100, 200$~GeV at $ \sqrt{s}=5.5$~TeV central
Pb+Pb collisions at the LHC. } 
\label{table:mean-radii}
\end{center}
\end{table}

\begin{figure}[!t]
\includegraphics[width=2.6in]{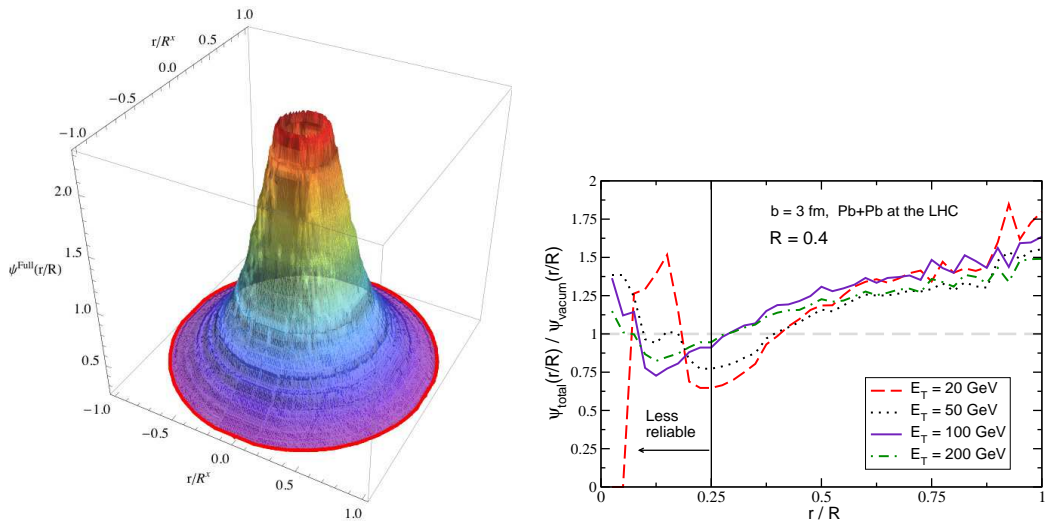}\hspace*{.3cm}
\includegraphics[width=2.6in]{Rpsi-R0.4-W0-jet-N.eps}
\vspace*{.1cm}
\caption{ Simulation of a jet shape in central Pb+Pb collisions at the LHC (left panel). 
The ratios of the medium-modified jet shape in heavy-ion
collisions to the jet shape in the vacuum for  jet energies
$E_T=20,\, 50,\, 100,\, 200$~GeV  for $R=0.4$ (left panel).
}
\label{fig:Rpsi-ratio}
\end{figure}

Lastly, we point out where the anticipated jet broadening effects will be
observed in the differential shape by studying  the ratio
$\psi_{\rm tot.}(r/R)/\psi_{\rm vac.}(r/R) $ in Figure~\ref{fig:Rpsi-ratio}.
The left panel shows a simulation of the differential shape of a 100~GeV
jet at the LHC. The right panel presents the ratio discussed above.  
We recall that the small $r/R < 0.25$ region of the intra-jet energy flow
in p+p collisions in our calculation has uncertainties associated
with the normalization of the jet shape. In the moderate and large $r/R>0.25$
region our theoretical  model gives excellent descriptions of the
Fermilab  Run II (CDF II) data. The QGP effects are manifested   as 
a suppression near the core and enhancement near the periphery of the jet.
In the tail of the energy flow distribution  for a cone radius $R=0.4$
this ratio can reach values  $\sim 1.75$.

\section{Conclusions}

In summary, we discussed the discriminating power of jet observables with respect to models of parton
propagation and energy loss in the QGP. More stringent constraints on theory, such as the ones that 
upcoming data on jet production and modification in heavy-ion reactions will provide, are necessary 
for precision jet tomography  and accurate determination of the properties of strongly-interacting 
nuclear matter. To this end, we presented perturbative QCD
calculations of inclusive jet cross sections/shapes and $Z^0/\gamma^*$-tagged jets to
${\cal O}(\alpha_s^3 )$, ${\cal O}(\alpha_s^2\alpha_s^{\rm med.} )$
and ${\cal O}(G_F\alpha_s^2 )$, ${\cal O}(G_F\alpha_s\alpha_s^{\rm med.} )$,
respectively. To demonstrate the sensitivity of these  observables to the characteristics of the  vacuum 
and  medium-induced parton showers, we examined their pronounced dependence on the 
cone radius $R$ at next-to-leading order. Experimental jet physics results
in heavy-ion collisions are still very preliminary but agree qualitatively with the theoretical 
expectations. Future studies in this direction will focus on the inclusion of non-perturbative hadronization effects and
extending the NLO calculations of jets at RHIC and at the LHC. In their entirety, these theoretical advances
will provide  first-principles insights into the many-body QCD parton dynamics at 
ultra-relativistic energies.

\section*{Acknowledgements}
I would like to thank KITP for their hospitality and support in the preparation of this brief overview article and  
my collaborators R.~B.~Neufeld and B.~W.~Zhang for helpful discussion. 
This research is also supported by the US Department of Energy, Office  of Science, under Contract No. DE-AC52-06NA25396 
and  by the LDRD program at LANL.

%

\end{document}